\documentclass[aps,prb,twocolumn,groupedaddress,eqsecnum,showpacs]{revtex4} 
\usepackage{amsfonts}
\usepackage{dcolumn}
\usepackage{graphicx,amssymb,amsmath}
\usepackage{bm}
\usepackage{natbib}
\usepackage{epstopdf}
\begin{document}

\title{Negative effective mass transition and anomalous transport in power-law hopping bands}

\author{Shimul Akhanjee}
\email[]{shimul@physics.ucla.edu}

\affiliation{Department of Physics, UCLA, Box 951547, Los Angeles, CA 90095-1547}


\date{\today}

\begin{abstract}

We study the stability of spinless Fermions with power law hopping $H_{ij} \propto \left|i - j\right|^{-\alpha}$. It is  shown that at precisely $\alpha_c =2$, the dispersive inflection point coalesces with the band minimum and the charge carriers exhibit a transition into negative effective mass regime, $m_\alpha^* < 0$ characterized by retarded transport in the presence of an electric field. Moreover, bands with $\alpha < 2$ must be accompanied by counter-carriers with $m_\alpha^* > 0$, having a positive band curvature, thus stabilizing the system in order to maintain equilibrium conditions and a proper electrical response. We further examine the semi-classical transport and response properties, finding an infrared divergent conductivity for 1/r hopping($\alpha =1$). The analysis is generalized to regular lattices in dimensions $d$ = 1, 2, and 3.
 
\end{abstract}


\maketitle

\section{introduction}

Conducting models with long-ranged, power law tight-binding bands(PLTB) have been introduced in a number of different physical scenarios such as in DNA chains\cite{nature}, Frenkel excitons\cite{dominguezprl}, and resonating valence bond (RVB) phases\cite{kuramotoPRL}. Recent theoretical work on PLTB bands has focused on disorder-induced critical phenomena, where critical delocalization occurs at $\alpha =1$ for random hopping\cite{mirlinpre1996,levitov} and a mobility edge in the regime $1<\alpha<3/2$ for non-random hopping and diagonal disorder\cite{dominguezprl}. Additionally, others have investigated variational magnetic states in strongly correlated Hubbard models at large onsite repulsion $U>>t$ resulting in the $t-J$ approximation with both long-ranged exchange $J_{ij}\propto \left|i - j\right|^{-2}$ and hopping\cite{shastryPRL, haldanePRL, kuramotoPRL} $t_{ij}\propto \left|i - j\right|^{-2}$. Earlier numerical investigations of non-disordered PLTB systems by Borland and Menchero\cite{borland_brazil} explored non-extensive effects of PLTB bands. They found that for finite chains of length $N$, for $0 <\alpha < 1$, there is anomalous wave-packet spreading and for $\alpha > 3/2$, ballistic motion is recovered as the nearest-neighbor limit is approached. However, there have been no systematic studies of the dispersion relations, or the transport properties of the non-disordered free-field model.

In this article we study various properties of PLTB chains as a function of the parameter $\alpha$. Here we restrict our analysis to those particular cases of $\alpha$ that are tractable analytically, resulting in simple closed form  representations of the dispersion relations. The effective mass is examined as a function of $\alpha$ and it shown precisely that at $\alpha_c=2$, the system enters into a negative effective mass regime, characterized by retarded transport in the presence of an electric field. Mathematically, this occurs at a critical value $\alpha_c$ where the dispersive inflection point moves to the bottom of the band until the long wavelength behavior is dominated by a negative curvature. For uncoupled isotropic tight-binding bands, our results and methods can be generalized to regular lattices in dimensions 1,2, and 3.

First let us clarify as to which energy regimes these band curvature effects play a significant role. The two primary limits to consider are near $k \to 0$ and $k=k_F$. The former case is near the bottom of the band where the concavity and the functional form of the dispersion relation will govern the semi-classical dynamics and transport in the presence of electric and magnetic fields. However, in the latter case these curvature effects are irrelevant near the Fermi surface, where any dispersion can be expanded as follows (in 1D for example),
\begin{equation}
\varepsilon (\vec k) = \mu  + \vec v_F  \cdot (\vec k - \vec k_F ) + \left. {\frac{{\partial ^2 \varepsilon }}{{\partial k^2 }}} \right|_{\vec k = \vec k_F } (\vec k - \vec k_F )^2 
\end{equation}
and the linear contribution is usually retained as $\varepsilon(k) \simeq v_F(k - k_F) $ near the Fermi points $\pm k_F$ without any loss of generality. Therefore, irrespective of the curvature away from $\varepsilon_F$, the low-energy Fermi-surface instabilites should remain unaffected as long as the dimensionality and Fermi-surface topology remains unchanged.

\section{The non-interacting band structure}
\subsection{The dispersion relations}

Consider the tight-binding model in d=1 (the analysis to higher dimensional isotropic lattices) with one orbital per site, for $N$ sites with lattice spacing $a$, governed by the following Hamiltonian (neglecting spin),
\begin{equation}
\mathcal{H}_\alpha = -\sum\limits_{i \ne j} {t_{ij} ^\alpha \psi_i^\dag  \psi_j  + h.c.}
\label{eq:hamilt}
\end{equation}
where $\psi_i^\dag$ and $\psi_j$ are fermion fields that obey $\left\{ {\psi _i^\dag  ,\psi _j } \right\} = \delta _{ji} $
. The effective width of the spectrum $t_{ij}$ has the power law spatial dependence,
\begin{equation}
t_{ij} ^\alpha = \frac{{t_0 }}{{\left| i-j \right|^\alpha  }}
\label{eq:hopping}
\end{equation}
where the constant $t_0$ depends on the details of the atomic orbital overlap matrix elements. For the entirety of this article we shall focus on the specific even positive integers $\alpha = 1, 2$ and $4$, given that the full dispersion relations can be treated analytically, rather than only retaining asymptotic forms near the top and bottom of the band. Moreover, this particular range of exponents encompasses the earlier ranges of observed criticality in disordered systems.
Because $t_{ij} ^\alpha \to 0$ as $\left| i-j \right| \to \infty $, a proper convergence of the sums taken for a large chain is ensured. For translationally invariant systems with periodic boundary conditions we can make use of the Bloch wave representation,
\begin{eqnarray}
\psi_k = \frac{1}{{\sqrt N }}\sum\limits_n {e^{ikn} \psi_n }  \\ 
 \mathcal{H}_\alpha = \sum\limits_k {\varepsilon_{k,\alpha} \psi_k^\dag \psi_k}  
\label{eq:bloch}
\end{eqnarray}
where the $k$'s run over the first Brillouin zone. In the limit of a large system size the band dispersion takes the form,
\begin{equation}
\begin{aligned}
\varepsilon_{k,\alpha }  &=  - t_0\sum\limits_{n = 1}^\infty  {\sum\limits_{z =  \pm n} {\frac{{e^{ikz} }}{{\left| n \right|^\alpha  }}} } \\
&= - t_0 \left( Li_\alpha \left[e^{ik}\right] + Li_\alpha \left[e^{-ik}\right] \right)
\end{aligned}
\label{disp1}
\end{equation}
where $Li_\alpha (z)$ is the polylogarithm function\cite{stegun}. For the aforementioned cases of $\alpha$, one can make use of the following exactly summable series, which are valid for $\left|k \right|< 2\pi$, which is beyond the natural cutoff of the 1st Brillouin zone\cite{stegun}(Note that we have absorbed a factor of 2 into $t_0$):

\begin{figure}
\centerline{\includegraphics[height=2.5in]{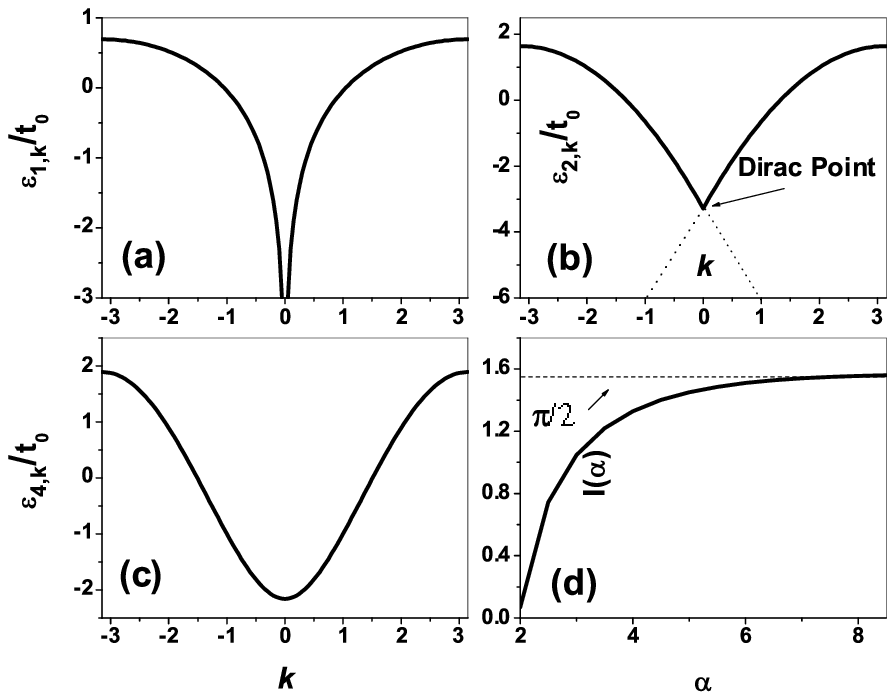}}
\caption{A comparison of the kinetic dispersion relations at different hopping powers of $\alpha$ ($\hbar=1$).(a) At $\alpha =1$, there is a logarithmic divergence at $k=0$. (b) For the case $\alpha =2$ corresponding to inverse-squared hopping, a Dirac cone emerges at smaller values of $k$ unlike the Tomonaga-Luttinger liquid, which is linearized around the two Fermi points $\pm k_F$, rather than at $k \to 0$. (c) At shorter ranged hopping, the conventional parabolic band is restored, with some additional curvature. (d) Numerical determination of the inflection point $I(\alpha)$ as a function of $\alpha$, where at precisely $\alpha_c =2$, $I(\alpha_c)=0$ (summations taken for N=500). }
\label{fig:disp}
\end{figure}

\begin{eqnarray}
\label{eq:disps1} &&\varepsilon_{k,1}  = t_0 \ln \left( {2\sin \left( {\frac{\left| k \right|}{2}} \right)} \right) \\  
\label{eq:disps2} &&\varepsilon_{k,2}  = - t_0 \left( {\frac{{\pi ^2 }}{3} - \pi \left| k \right|
 + \frac{{k^2 }}{2}} \right) \\  
\label{eq:disps3} &&\varepsilon_{k,4}  = - t_0 \left( {\frac{{\pi ^4 }}{{45}} - \frac{{\pi ^2 k^2 }}{{6}} + \frac{{\pi \left| k \right|^3 }}{{6}} - \frac{{k^4 }}{{24}}} \right)
\end{eqnarray}
The Eqs.({\ref{eq:disps1}-\ref{eq:disps3}) have been plotted in Fig.\ref{fig:disp}. Apparently, the preceding expressions produce the full band curvature within the first Brillouin zone, however in certain cases the explicit periodicity is absent but can be repeated to construct higher Brillouin zones. Time reversal invariance is preserved such that $\varepsilon_{k,\alpha}=\varepsilon_{-k,\alpha}$. Note that the analysis can generalized to all isotropic regular lattices by simply reproducing each dispersion independently in each direction. For example the $\alpha =1$ case in $d=3$ cubic lattice can be written as,
\begin{equation}
\varepsilon _{k,1}^{d = 3}  = t_0 \ln \left[ {8\sin \frac{{\left| {k_x } \right|}}{2}\sin \frac{{\left| {k_y } \right|}}{2}\sin \frac{{\left| {k_z } \right|}}{2}} \right]
\end{equation}

Apparently, the range of the interactions has a noticeable effect on the precise functional form of the dispersion relations, near $k \to 0$. The first and second derivatives of  $\varepsilon_{k,\alpha}$ respectively give rise to the group velocity $v_\alpha (k) = \frac{1}{\hbar }\frac{{\partial \varepsilon_{ k,\alpha}}}{{\partial k}}$ and the effective mass $m_\alpha ^* (k) = \hbar ^2 \left( {\frac{{\partial ^2 \varepsilon_{k,\alpha}}}{{\partial k^2 }}} \right)^{ - 1}$. Taking the appropriate derivatives we have for the group velocity,
\begin{equation}
\begin{array}{l}
 v_1 (k) =   \frac{{t_0 }}{\hbar }\cot (\left| k \right|/2) \\ 
 v_2 (k) =  - \frac{{t_0 }}{\hbar }(\left| k \right| - \pi ) \\ 
 v_4 (k) =  - \frac{{t_0 }}{\hbar }\left( { - \frac{{\pi ^2 {\rm{\left| k \right|}}}}{{\rm{3}}}{\rm{ + }}\frac{{\pi {\rm{\left| k \right|}}^{\rm{2}} }}{{\rm{2}}} - \frac{{{\rm{\left| k \right|}}^{\rm{3}} }}{{\rm{6}}}} \right) \\ 
 \end{array}
 \label{eq:vel}
\end{equation}
and the effective mass,
\begin{equation}
\begin{array}{l}
 m_1^* (k) = -\frac{{2\hbar ^2 }}{{m_e t_0 }}\sin ^2 \left( {\left| k \right|/2} \right) \\ 
 m_2^* (k) =  - \frac{{\hbar ^2 }}{{m_e t_0 }} \\ 
 m_4^* (k) =  - \frac{{2\hbar ^2 }}{{m_e t_0 \left( { - \frac{{\pi ^2 }}{{\rm{3}}}{\rm{ + }}\pi {\rm{\left| k \right|}} - \frac{{{\rm{\left| k \right|}}^{\rm{2}} }}{{\rm{2}}}} \right)}} \\ 
 \end{array}
 \label{eq:mass}
\end{equation}
which have been plotted in Fig.\ref{fig:vel}. It follows that an inflection point $I(\alpha)$ can be defined as the location where $m_\alpha ^* (k)=0$.

\begin{figure}
\centerline{\includegraphics[height=2.5in]{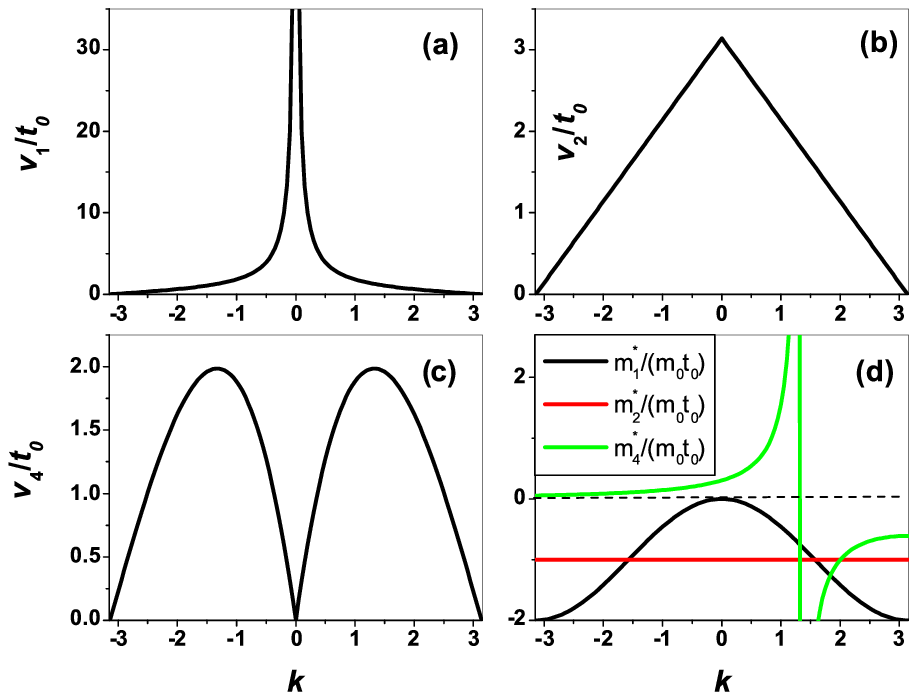}}
\caption{The group velocity and effective mass of the electrons ($\hbar=1$) (a) $\alpha=1$, the velocity diverges as the effective mass tends to zero for small $k$. (b) At $\alpha=2$ the velocity steadily decreases for a constant negative effective mass. (c) $\alpha=4$ is relatively short-ranged, with a velocity that increases then decreases, with a shift to negative effective mass values originating from an inflection point in the dispersion. (d) The effective mass for each case of $\alpha$. Notice the emergence of negative values at $\alpha = 1,2$ for small k, where the inflection point has disappeared, and the curvature of the dispersions becomes purely negative.}
\label{fig:vel}
\end{figure}

\section{The negative effective mass regime }

\subsection{Semi-classical dynamics }

In the usual particle/hole theory, electrons with charge $e < 0$ are accelerated from the bottom of a parabolic band when an external positive electric field is applied. Electrons are therefore defined to have a positive effective mass, $m_\alpha^* > 0$. On the other hand, holes possess the opposite charge $e > 0$ and move in a (valence) band, with a negative curvature, generating a negative effective mass $m_\alpha^* < 0$ and thus will move in the opposite direction as the electron. Hence, the concavity and curvature of a band, will govern the sign and magnitude of $m_\alpha^* > 0$ as a function of $k$. Semi-classically, the Boltzmann equation states that the fields $\vec E$ and $\vec H$ will change the local carrier concentration $f_k$ at the rate\cite{zimansolid},
\begin{equation}
\begin{aligned}
 \left. {\frac{{\partial f_k }}{{\partial t}}} \right]_{field}  &=  - \frac{{d\vec k}}{{dt}} \cdot \frac{{\partial f_{\vec k} }}{{\partial \vec k}} \\ 
  &=  - \frac{e}{\hbar }\left( {\vec E + \frac{1}{c}\vec \nabla \varepsilon _{\vec k,\alpha }  \times \vec H} \right) \cdot \frac{{\partial f_{\vec k} }}{{\partial \vec k}} \\ 
 \end{aligned}
\end{equation}
Consequently, electrons and holes, with reversed dispersive concavity do not have the conventional metallic response to an applied electric field; there is no acceleration of the carriers, rather the opposite occurs and the electronic transport is suppressed by a current $I_{<}$\cite{smolin}. If such carriers participate in the conduction process, then a positive current recorded by the measuring instrument must be accompanied by an internal current $I_o = I_{<}+I_{>}  $, where $I_{>}$  is part of the positive current due to particles having $m_\alpha^* > 0$. Thus, multiple band with both types of carriers should be introduced, such that a net positive effective mass can maintain equilibrium.

\subsection{Transition at $\alpha = 2$}

For a large range of $\alpha$'s considered, there is a definite transition from a positive to negative band curvature.
By only considering the exact expression given by Eqs.(\ref{eq:mass}) it is clear that $2 \le \alpha_c <4 $.
Take the extreme case of short ranged hopping, which is the nearest-neighbor limit and can be approached at large $\alpha$, and the dispersion becomes $\varepsilon_{nn}(k) =-2t_0 \cos k$ with $m_\alpha^* > 0$ near the band minimum, having an inflection point exactly at $I_{nn}=\pm \pi /2$. The other limit of interest, can be examined more precisely by performing the summations of Eq.(\ref{disp1}) numerically as shown in Fig.\ref{fig:disp} (d). Notice that as $\alpha$ becomes smaller, the inflection point moves closer to the band minimum, $I(\alpha_c)=0$ until the curvature has reversed sign, at $\alpha \approx 2$. Because the summations converge rapidly the finite size effects are negligible. Additionally, for a more precise analysis, one can make use of the derivative relation $\frac{{\partial Li_s  (e^\mu  )}}{{\partial \mu }} = Li_{s  - 1} (e^\mu  )$. Consequently, the effective mass is controlled by the function $Li_s(z)$ at negative values of $s$ when $\alpha < 2$ and in order to discern the behavior of the $Li_s(z)$ function at negative indices, we can employ the expansion\cite{stegun}, 
\begin{equation}
Li_{ - s} (z) = \frac{1}{{(1 - z)^{m + 1} }}\sum\limits_{k = 1}^m {a_{m,k} z^k } {\kern 1pt} {\kern 1pt} {\kern 1pt} {\kern 1pt} {\kern 1pt} {\kern 1pt} {\kern 1pt} {\kern 1pt} {\kern 1pt} {\kern 1pt} {\kern 1pt} {\kern 1pt} {\kern 1pt} {\kern 1pt} {\kern 1pt} {\kern 1pt} {\kern 1pt} {\kern 1pt} {\kern 1pt} {\kern 1pt} {\kern 1pt} {\kern 1pt} {\kern 1pt} {\kern 1pt} {\kern 1pt} {\kern 1pt} (s > 0)
\end{equation}
where the coefficients are Eulerian numbers that satisfy the recurrence relations,
\begin{equation}
a_{m,k}  = (m + 1 - k)a_{m - 1,k - 1}  + ka_{m - 1,k} 
\end{equation}
Taking the appropriate form of Eq.(\ref{disp1}), it is clear that $m_\alpha^*< 0$ in the range $0 < \alpha < 2$.

\subsection{Transport in 1/r bands} 

For the longest ranged hopping exponent considered ($\alpha =1$), there is a unique logarithmic divergence at long wavelength values of $k$, which is accompanied by a vanishing effective mass, as shown in Figs. \ref{fig:disp}(a) and \ref{fig:vel}(a), and with the reduced inertia the velocity is divergent. Such a singularity will have important consequences and requires a careful interpretation if such an equilibrium state is valid. The semi-classical conductivity tensor in the relaxation-time approximation is given by\cite{zimansolid},
\begin{equation}
\begin{aligned}
 \sigma^{nn'} _\alpha   = \frac{2}{{(2\pi )^d }}\frac{{e^2 \tau (\varepsilon _F )}}{\hbar }\int_{Fermi} {\frac{{ v_\alpha ^n (k) v_\alpha ^{n'} (k)d^d S_F }}{{v_\alpha  (k)}}}  \\ 
  = \frac{2}{{(2\pi )^d }}\frac{{e^2 \tau (\varepsilon _F )}}{\hbar }\int_{Fermi} {\mathbf{M}^{ - 1} d^d S_F }  \\ 
 \end{aligned}
\end{equation}
where the integrations are taken over the Fermi surface, $\tau (\varepsilon _F )$ is the scattering lifetime at the Fermi level and $\mathbf{M}$ is the effective mass tensor. For systems with cubic symmetry at arbitrary filling, the tensor reduces to a scalar yielding
\begin{equation}
\begin{aligned}
 \sigma _\alpha   &= \frac{2}{{(2\pi )^d }}\frac{{e^2 \tau (\varepsilon _F )}}{\hbar }\int_{Fermi} {\left| {\vec v_\alpha  (k)} \right|d^d S_F }  \\ 
  &=  - \infty {\kern 1pt} {\kern 1pt} {\kern 1pt} {\kern 1pt} for{\kern 1pt} {\kern 1pt} {\kern 1pt} {\kern 1pt} d = 1,2,3 \\ 
 \end{aligned}
\end{equation}
which is logarithmically divergent at the infrared limit, which coincides with the previously observed anomalous wavepacket spreading\cite{borland_brazil}, now also exhibiting anomalous transport characteristics. Another point to consider is the behavior of the Fermi level $E_F$ as a function of the filling fraction or density. At $\alpha =1$ the Fermi energy depends on the density $\rho$ as,
\begin{equation}
E_F^1 (r_s ) =  t\ln \left( {2\sin \frac{\pi\rho }{{2 }}} \right)
\label{eq:1fermi}
\end{equation}
Upon inspection,  Eq.(\ref{eq:1fermi}) vanishes at $\rho = 1/3$ or a one-third filling fraction unlike the nearest-neighbor band which has a Fermi energy that vanishes at half filling or $\rho = 1/2$.  In general charge/spin density wave formation, superconductivity and Mott metal-insulator transitions are strongly filling fraction dependent. Therefore, it would be of interest to determine if there are $\rho = 1/3$ band instabilities in the system in the presence of interactions, as a negatively diverging Fermi energy indicates that the $\alpha =1$ system is unstable at lower densities.

\subsection{$1/r^2$ hopping}

Of much interest is close to the transition point of $\alpha =2$, or inverse squared hopping. It follows that at small $k$, the dispersion is linear, $\varepsilon_{k,2} \propto k$, which maps onto a pseudo-relativistic kinetic energy, analogous to the 6 Dirac points of graphene. However, unlike the Tomonaga-Luttinger liquid, $\varepsilon_{k,2}$ is not linearized close to the two Fermi points $\pm k_F$ and does not require a cutoff in order to remedy negative energy states\cite{quantum1D}. It should be noted that other models that are relevant to $\alpha =2$ hopping were studied by Haldane\cite{haldanePRL} and Shastry\cite{shastryPRL}, who determined the ground state of the spin 1/2 antiferromagnetic Heisenberg chain with $1/r^2$ exchange. The Haldane-Shastry model was exactly solved by a Gutzwiller projected wavefunction, identifying this particular class of inverse squared exchange models with Anderson's RVB phase\cite{andersonrvb}. For a PLTB band the exchange constant $J$, in the limit of large repulsion $U>>t$, is given by 
\begin{equation}
J_{ij}^\alpha   = \frac{{4\left( {t_{ij}^\alpha  } \right)^2 }}{U}
\end{equation}
Thus, the point here to emphasize is that long range exchange and hopping are related, as it requires a $1/r$ kinetic band to produce a $1/r^2$ exchange constant and therefore the earlier $1/r$ and negative effective mass transport characteristics become obviously important. Later, Kuramoto \emph{et al}\cite{kuramotoPRL} have included inverse squared hopping in addition to inverse-squared exchange, in a supersymmetric $t-J$ model, showing that it is also exactly solved by a Gutzwiller projected wavefunction. Hence, the $\alpha=1,2$ cases with negative effective mass carriers are relevant to particular classes of RVB phases and strongly correlated systems. The consequences of this apparent connection needs to be explored further as the long-ranged hopping appears to control the magnetic frustration.

\subsection{Bulk properties for $\alpha \le 2$, d=1}
Let us further investigate the properties of the negative effective mass regime in 1D. Another method for probing the stability of the ground-state is to study the total energy. The emergence of a complex or imaginary component of the thermodynamic ground-state energy suggests that the system is in a non-equilibrium state. The density for a degenerate Fermi system is $\rho = k_F / \pi$ and the total kinetic energy per particle can be obtained by integrating over the internal energy distribution,
\begin{equation}
U_\alpha (\rho) = \frac{1}{N} \int\limits_0^{E_F } {\lambda g_\alpha  (\lambda)} d\lambda 
\label{eq:internal}
\end{equation}
The density of states can by easily computed by inverting the dispersion relations in terms of $k$ and using the relation $g(\lambda)=2/\pi \hbar v(\lambda )$. After substituting the velocity expressions from (\ref{eq:vel}), we have 

\begin{eqnarray}
\label{eq:dos1} g_1 (\lambda ) &=& \frac{{2e^{\lambda /t_0 } }}{{\pi t_0 \sqrt {1 - (1/4)e^{2\lambda /t_0 } } }} \\ 
\label{eq:dos2} g_2 (\lambda ) &=& \frac{{2\sqrt 3 }}{{\pi \sqrt {t_0 (\pi ^2 t_0  - 6\lambda )} }}  
\end{eqnarray}
Subsequently, deep in the negative effective mass regime, for $\alpha=1$, the total energy becomes,
\begin{equation}
\begin{aligned}
 &U_1 (\rho ) = \int\limits_{\ln 2}^{E_F^1 } {\lambda g_1 (\lambda )d\lambda }  \\ 
  &= 2t_0 \left( {\frac{{i\pi ^2 \rho ^2 }}{4} + \pi \rho \ln \left[ { - ie^{i\pi \rho /2} } \right] - iLi_2 \left[ { - e^{ - i\pi \rho } } \right]} \right) + C^1  \\ 
 \end{aligned}
 \label{eq:u1}
\end{equation}
where $C^1$ is a complex constant. $U_1$ is purely real as a function of density in the range $(0,1)$, indicating that there are no obvious bulk instabilities. Next let us examine the ground-state energy at $\alpha=2$, which is in the critical regime. After substituting Eq.(\ref{eq:dos2}) into Eq.(\ref{eq:internal}), we have the following density $\rho$ dependent ground-state energy
\begin{equation}
\begin{aligned}
 U_2 (\rho ) &= \int\limits_{\pi ^2 /6}^{E_F^2 } {\lambda g_2 (\lambda )d\lambda }  \\ 
  &= \frac{{\pi ^2 t_0 }}{6}\left( {2\rho (2 + \rho (\rho  - 3)) - 1} \right) \\ 
 \end{aligned}
 \label{eq:u2}
\end{equation}
which is purely real, indicating a stable ground-state along with a stable Fermi energy. Lastly, the low energy scaling in terms of the density is often used to compare with Coulomb interactions or disorder. The well-known competition between the kinetic(band) energy and other terms can result in a change of the ground-state. Therefore, a naive low energy scaling of the total energies results in the following,
\begin{eqnarray}
 U_1(\rho)&\propto& \ln (\rho ) \\ 
 U_2(\rho)&\propto& \rho  
 \end{eqnarray}

\section{Fermi surface nesting and screening at $\alpha =2$}

Evidently, carriers with $m_\alpha^* < 0$, will screen a charged impurity differently from the free Fermi gas. Therefore it would be useful to examine the susceptibility of the particle-hole channel to highlight some of these differences. 
Lindhard theory is an important perturbative approach to quantify the change in the electronic charge density $\delta n $ due to a static impurity potential $\phi ^{ext}$. In Appendix \ref{appena} we present the conventional formalism used for the 1D free Fermi gas. Here, we shall focus on the analytically tractable 1D, T=0 behavior of the static susceptibility $\chi _0^{2} (q)$ at the negative mass transition($\alpha =2$). Subsituting the form for $\varepsilon_{k,2}$ into Eq.(\ref{eq:lind}) yields the following,
\begin{equation}
\begin{aligned}
\chi _0^2 (q)&=\int\limits_{ - k_F }^{k_F } {\left[ {\frac{{dk}}{{\varepsilon _{k + q,2}  - \varepsilon _{k,2} }} + \frac{{dk}}{{\varepsilon _{k - q,2}-\varepsilon _{k,2} }}} \right]}  \\ 
&=\frac{1}{{qt_0 }}\ln \left[ {\frac{{(q - (2k_F  + 2\pi ))(q - (2k_F  - 2\pi ))}}{{(q + (2k_F  - 2\pi ))(q + (2k_F  + 2\pi ))}}} \right]
\end{aligned}
\label{eq:lind2}
\end{equation}
which unlike the parabolic band or nearest neighbor case, does not exhibit a nesting singularity exactly at $q=2k_F$, rather it is shifted to $q=\pm 2(\pi - k_F )$. It follows that we can restore the lattice constant and it becomes clear that the nesting is lattice dependent, 
as the spacing of the reciprocal lattice determines the nesting,
\begin{equation}
\left| q \right|= \left| {2k_F  \pm \left| {\vec Q} \right|} \right|
\label{eq:nest}
\end{equation}
This dependence has futher implications for the RKKY function, which for a delta function impurity potential is simply the Fourier transform  Eq.(\ref{eq:lind2}),
\begin{equation}
\delta _2 n(x) = \frac{{ - \sqrt {2\pi }}}{{t_0 }}\left[ {si\left( {(\left| {\vec Q} \right| - 2k_F )x} \right) - si\left( {(\left| {\vec Q} \right| + 2k_F )x} \right)} \right]
\label{eq:rkky2}
\end{equation}
Consequently the at large distances we have,
\begin{equation}
\begin{aligned}
 \delta _2 n(x) &\propto \frac{{ - 1 }}{{t_0 x}}\left[ {\cos \left( {(\left| {\vec Q} \right| - 2k_F )x} \right) - \cos \left( {(\left| {\vec Q} \right| + 2k_F )x} \right)} \right] \\ 
  &\propto  -  \frac{{\cos ( {\left| {\vec Q} \right|x} )\sin \left( {2k_F x} \right)}}{{t_0 x}} \\ 
 \end{aligned}
 \label{eq:friedel2}
\end{equation}
where Eq.(\ref{eq:friedel2}) exhibits interference fringes(or the mathematical form of beating) instead the usual case of purely oscillatory behavior shown in Eq.(\ref{eq:friedelnn}). Moreover, the condition for constructive interference requires that $(\left| {\vec Q} \right| - 2k_F )x = 2\pi n$ for $n = 0,1,2...$. This suggests that in inverse-squared hopping conducting materials, the umklapp momenta can be derived from the interference fringe maxima of the Friedel oscillations. We can further examine the special case of the half-filled band where $\left| {\vec Q} \right| = 4k_F$. This leads to 
\begin{equation}
\delta _2 n(x) \propto  - \frac{{\cos \left( {4k_F x} \right)\sin \left( {2k_F x} \right)}}{{t_0 x}}
\end{equation}
which contains unusual $4k_F$ periodicity, reminiscent of the 1D Wigner crystal density-density oscillations\cite{quantum1D}.

\section{Conclusion}
In summary we have explored various properties and the general dependence of the band curvature on the hopping exponent $\alpha$ for power-law banded electron systems. Our methods and approach can be generalized to alternative lattices in higher dimensions, and this article should serve as a basis to do so. We have demonstrated, using a variety of approaches, that the curvature completely changes sign in the range $\alpha_c = 2$ and have highlighted various static properties of the $\alpha=2$ system, including unusual lattice dependent Fermi surface nesting properties. In addition, our investigations are particularly useful for those studying Anderson localization and critical phenomena in Hamiltonians that contain long-ranged band structures as previous studies on such systems  have highlighted range $1< \alpha < 1.5 $ as a critical regime. Here have demonstrated that even in the absence of disorder, the  $\alpha =1$ system is not stable at lower densities and contains logarithmic divergence of its Fermi energy and anomalous transport characteristics.  

In closing we emphasize that the curvature of a band is generally associated with the magnitude and sign of the effective mass, resulting in important implications for electronic devices that require non-equilibrium conditions. For example, it has been shown that for negatively charged carriers with a negative effective mass, the current density is negative and the both the Hall emf and the thermal emf are positive, which follows directly from the Lorentz force law\cite{smolin}. Therefore, the capability of tuning a conducting system into a negative effective mass regime could be exploited for technological purposes, given that the interplay of non-equilibrium conditions with disorder and quantum/thermal fluctuations can affect the criticality and the transport properties. Additionally, an unexplored connection with strongly correlated models utilizing PLTB bands leading to magnetically frustrated spin-liquid states will be explored in a future work.

\begin{acknowledgments}
I would like to give thanks to Prof. Joseph Rudnick, for useful discussions and assistance. Also I would like to acknowledge F. Dominguez-Adame and V.A. Malyshev for pointing out several references. This work was supported by UC General Funds: 4-404024-RJ-19933-02
\end{acknowledgments}
\appendix
\section{Static response of a 1D free Fermi gas}
\label{appena}
 The general dielectric response $\epsilon (q) $ can be evaluated from the following linear response relation,
\begin{equation}
\epsilon (q) = \frac{{\phi ^{ext} (q)}}{{\phi (q)}} = 1- \phi ^{ext} (q)\chi (q)
\end{equation} 
where $\phi (q)$ is the full physical potential in $q$ space and the static Lindhard response function in 1D is given by,
\begin{equation}
\chi _0^\alpha  (q) = 2\mathcal{P}\int {\frac{{dk}}{{2\pi }}} \frac{{f_{k + q}  - f_k }}{{\varepsilon_{k + q,\alpha }  - \varepsilon_{k,\alpha } }}
\label{eq:lind}
\end{equation}
where $\mathcal{P}$ denotes the principal part of the integral and $f_k$ is the Fermi-Dirac distribution.
For the conventional 1D non-interacting electron gas at $T=0$, this evaluates to 
\begin{equation}
\chi _0^{nn} (q) = \frac{{2m}}{{\pi \hbar ^2 q}}\ln \left| {\frac{{2k_F  + q}}{{2k_F  - q}}} \right|
\label{eq:chinn}
\end{equation}
which has the well known singularity at $q = 2 k_F$, associated with Fermi surface nesting and spin/charge density wave instabilities. 
Furthermore, regarding the screening behavior, the simplest case assumes a delta function impurity potential of the form,
\begin{equation}
\phi ^{ext} (x) = \frac{{\hbar ^2 u_0 }}{{2m}}\delta (x)
\label{eq:imppot}
\end{equation}
which leads to the following real space density modulation:
\begin{equation}
\delta _\alpha  n(x) = \frac{{\hbar ^2 }}{{2m}}\sum\limits_q {\chi _0^\alpha  (q)e^{iqx} } 
\label{eq:deltan}
\end{equation}
After subsitituting Eq.(\ref{eq:chinn}), one has the following expression
\begin{equation}
\delta  n_{nn}(x) =  - \frac{{u_0 }}{\pi }\int\limits_x^\infty  {\frac{{\sin t}}{t}} dt =  - \frac{{u_0 }}{\pi }si(2k_F x)
\end{equation}
which is the well known Ruderman-Kittel-Kasuya-Yosida (RKKY) range function. The large distance asymptotic becomes,
\begin{equation}
\delta n_{nn}(x)\propto - \frac{{\cos (2k_F x)}}{x}
\label{eq:friedelnn}
\end{equation}
displaying the $2k_F$ periodicity and $1/x$ decay commonly referred to as Friedel oscillations\cite{prbrkky}.

\end{document}